# Reservoir Computing-based Multi-Symbol Equalization for PAM 4 Short-reach Transmission

**Yevhenii Osadchuk, Ognjen Jovanovic, Darko Zibar, Francesco Da Ros**
*DTU Electro, Technical University of Denmark, DTU, Kongens Lyngby, 2800, Denmark*
*yevos@dtu.dk*

**Abstract:** We propose spectrum-sliced reservoir computer-based (RC) multi-symbol equalization for 32-GBd PAM4 transmission. RC with 17 symbols at the output achieves an order of magnitude reduction in multiplications/symbol versus single output case while maintaining simple training. © 2022 The Author(s)

## 1. Introduction

The ever-increasing demand for high data rates and low latency in short-reach metro and access networks propels the need for simple and low-cost infrastructure with a small footprint. The intensity-modulated directly detected (IM/DD) links are the primary candidates for such short-reach interconnects [1]. However, the IM/DD links with square-law photodetection (PD) convert chromatic dispersion (CD) into a nonlinear inter-symbol interference (ISI) accumulated during the propagation in fiber. To mitigate the ISI, recurrent neural networks (RNNs) with sliding window input have been proposed in [2], achieving substantial bit-error-rate (BER) performance. However, the RNNs incorporate high computational complexity (CC) in terms of real multiplications per equalized symbol (RMPS) and are challenging to train [3]. Therefore, reservoir computing (RC), a special type of RNN, has been introduced in [4,5] to compensate for fiber-induced impairments, while keeping computational and training complexity low. In [6], it was shown that introducing a sliding window at the input boosts the RC performance for time series prediction tasks. Usually, in NNs-based digital signal processing blocks, the equalization is performed sequentially symbol by symbol [7]. However, it was experimentally demonstrated in [3] that feedforward NN-based multi-symbol equalization achieved considerable CC relaxation without losing in BER.

In this work, we inherit the idea of using a sliding window at the input of the RC and propose a multi-symbol RC equalizer to decrease CC without impairing performance. We benchmark the BER performance and complexity to a single-symbol RC for 32 GBd PAM4 short-reach transmission by first dividing the signal`s spectrum into 4 parts and then detecting it with separate PDs. By feeding sliding window input into RC, we discover that equalizing in total up to 17 symbols at once significantly decreases CC without penalty in BER performance.

## 2. Multi-symbol RC-based Equalizer Design for PAM 4 IM/DD

The system under investigation is shown in Fig. 1. A random bits sequence is generated, and shaped with RRC (roll-off 0.1) at 2 samples per symbol (sps) to construct a 32 GBd PAM4 signal. The electrical signal modulates the optical carrier (laser) in a Mach-Zehnder Modulator (MZM) and is then guided into a single-mode fiber with a CD as a main limiting distortion (D=16.4 ps/nm/km). At the receiver, the spectrum of the signal is sliced into 4 equal portions as in [4], each detected by an independent square-law PD, and followed by an electrical amplifier modeled as an additive white Gaussian noise (AWGN) source. The sliced signal sequence is then fed into an RC-based equalizer.

We define the input to the RC equalizer as $N_{in} = M * sps * 4$, where memory $M = (2k + 1)$ is designed with a sliding window, taking into account $k$`s previous and forward symbols. The number of reservoir nodes $N_{res} = 30$ for the multi-symbol RC equalizer with the $k$ equal to 11, spectral radius equal to 1.2, and leaking rate equal to 0.7. The $s_{in}$, $s_{res}$, and $s_{out}$ are the sparsity coefficients for the input, reservoir, and output weights matrices, which are equal to 0.1, 0.05, and 0.1 correspondingly. The number of symbols at the reservoir`s output is $N_{out}$.

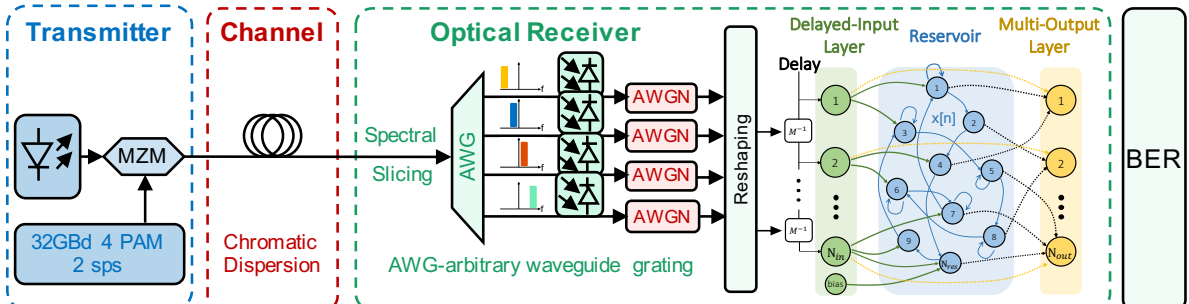

Fig. 1. PAM4 IM/DD transmission link with spectral slicing and RC-based equalizer at the receiver.

The described hyperparameters as well as the memory $M$ were obtained through Bayesian optimization [8] to achieve the best equalization performance. All the RCs are trained using 15% of $2^{22}$ symbols and tested on the rest of the data.

### 3. Results and Discussions

The BER versus signal-to-noise ratio (SNR) of RC equalizers with sliding window $M$ and $N_{out}$ symbols at the output (RC-($M$-to-$N_{out}$)), for $l = 10$ km transmission, is shown in Fig. 2a. The RC-(23-to-1) for $l$=0 km is added as a benchmark without ISI. The RC-based equalizer with a sliding window RC-(23-to-1) achieves better BER than a single-input RC-(1-to-1) with 300 reservoir nodes and optimized architecture adopted from [3] with and without spectrum slicing. We demonstrate that the BER performance of the optimized single-output RC-(23-to-1) can be achieved with a multi-symbol RC equalizer with up to 17 symbols at the output RC-(23-to-17). At the same time, the RC-(23-to-23) is unable to compensate ISI of the very first and last symbols in the input sequence.

In Fig. 2b, the SNR penalty at the KP4 FEC (BER=$2.26 \times 10^{-4}$) threshold against the RC-(23-to-1) $l = 0$ km versus the fiber length for RCs with a different number of output symbols is shown. We can see that the RC-(23-to-17) provides a similar to the best-optimized RC-(23-to-1) equalization gain up to 50 km transmission. Increasing only the output symbols from 17 up to the maximum of 23 significantly increases the SNR penalty due to lacking the neighboring memory of the symbols at the edges of the input sequence. Additionally, we can see that having an extra memory at the tails of the input window significantly improves the BER performance of an equalizer compared to the full stack RC-(23-to-23).

Fig. 2c shows the CC in terms of RMPS for different RC equalization types and it is calculated following the equation [9]: $CC = [N_{in}N_{res}s_{in} + N_{res}^2 s_{res} + N_{res}N_{out}s_{out} + 2N_{res} + N_{out}(1 + N_{res})]/N_{out}$. The RC-(23-to-17) can achieve the same equalization performance as the optimized RC-(23-to-1), having a magnitude reduction in complexity measure. This shows that the RC-based equalizer can mitigate CD-induced impairments with a multi-symbol approach significantly reducing the complexity.

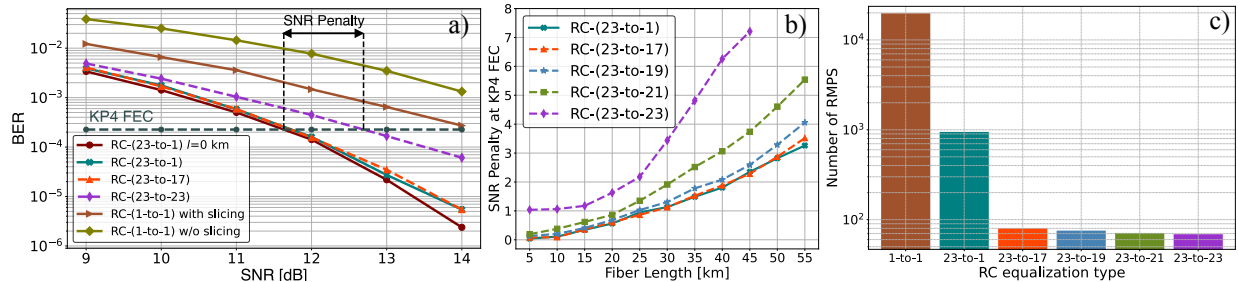

Fig. 2. (a) BER vs SNR RC equalization for 10 km transmission. (b) SNR Penalty at KP4 FEC threshold against the B2B vs the fiber length. (c) RMPS for RC-based equalizers with sliding window and RC-(1-to-1) with slicing.

### 4. Conclusions

The multi-symbol reservoir computing (RC)-based equalization has been numerically evaluated in a 32GBd PAM4 IM/DD transmission, showing superior performance compared to single-input approaches. In contrast to the best-performing RC equalizer with sliding window memory and a single output, the RC with 17 symbols at the output achieves an order of magnitude of complexity relaxation requiring only around 72 multiplications per equalized symbol, while keeping equivalent BER performance for up to 50 km transmission.

**Acknowledgments** Villum Fonden's YIP project OPTIC-AI project (grant no 29344) and ERC CoG FRECOM (grant no. 771878).